\documentstyle[epsfig,longtable]{aipproc}

\begin{document}
\title{Electroweak Sudakov corrections}

\author{Michael Melles$^*$}
\address{$^*$Paul Scherrer Institute, Villigen CH-5232, Switzerland}

\maketitle

\begin{abstract}
At energies much larger than the mass of the weak gauge bosons,
electroweak radiative corrections can lead to significant corrections. At 1 TeV
the one loop corrections can be of ${\cal O} \left( 20 \% \right)$ due
to large contributions of the Sudakov type. We summarize recent progress
in the evaluation and resummation of the double and single logarithmic
corrections to general scattering amplitudes for fermions, transversely
as well as longitudinally polarized external lines.
\end{abstract}

With the advent of future colliders in the TeV range (such as the LHC or
linear $e^\pm$ colliders), the evaluation of electroweak radiative corrections
has become an important ingredient in the search for new physics signals.
In principle one loop corrections can be calculated reliably in perturbation
theory and contain the largest part of the higher order corrections - up to
${\cal O} \left( 20 \% \right)$ at TeV energies. Precision measurements
at LEP have demonstrated the extraordinary success of the Standard
Model (SM) and thus restricted any extensions to a high degree of conformity,
i.e. all phenomenologically viable field theoretical models must not differ to a large degree
from the SM for $\sqrt{s}\sim M$. Thus it
is likely that the physics responsible for the spontaneous breaking of
the $SU(2)\times U(1)$ gauge symmetry -if different from the SM Higgs
mechanism- can only be disentangled from competing scenarios by measurements
in the percentile precision 
regime\footnote{We do not consider theories with large extra spatial
dimensions at the TeV scale for now.}. 
Typical scenarios at energies above the weak scale include establishing
supersymmetric relations between couplings, what type of supersymmetry would
be realized in nature, investigating the nature of possible strong
interaction physics in the $W^\pm$ sector, is the SM-Higgs sector minimal or establishing the
Higgs mechanism experimentally by reconstructing the potential.
Thus, also two loop electroweak corrections, typically in the percent-regime at
TeV energies, cannot be neglected.
While at LEP energies radiative corrections from QCD dominate Drell Yan processes
for instance, electroweak corrections can become comparable and even dominating
at high energies.  
In general one expects the
SM to be in the unbroken phase at high energies.
There are, however, some important differences of the electroweak theory with respect to an unbroken
gauge theory. Since the physical cutoff of the massive gauge bosons is the weak scale $M\equiv
M_{\rm W} \sim M_{\rm Z} \sim M_{\rm H}$,
pure virtual correction lead to physical cross sections depending on the infrared ``cutoff''.
This is the indeed the reason for the large size of the Sudakov corrections.
The photon needs to be treated in
a semi-inclusive way.
Additional complications arise due to the mixing involved to make the mass eigenstates and the fact
that at high energies, the longitudinal degrees of freedom are not suppressed.
Furthermore, since the asymptotic states are not group singlets, it is expected
that fully inclusive cross sections contain Bloch-Nordsieck violating electroweak corrections
\cite{ccc}.
In Ref. \cite{flmm} the leading QCD-like DL corrections were resummed to all orders using the
gauge invariant infrared evolution equation method \cite{kl} and a non-Abelian version
of Gribov's bremsstrahlung theorem \cite{vg}. 
The DL corrections for longitudinal degrees of freedom were obtained in
Ref. \cite{m1} via the Goldstone boson equivalence theorem. The kernel of the
equation differs in the high energy regime where the virtuality of the exchanged gauge bosons
is larger than the weak scale $M$ and the regime below, where only photonic corrections
need to be considered. The explicit two loop DL calculations in Refs. \cite{mgap,bw,hkk} with the physical
SM fields for the process $g \longrightarrow f {\overline f}$ have confirmed this picture.
In addition, the universal subleading corrections for external fermions, transversely as
well as longitudinally polarized gauge bosons from Refs. \cite{m1,m2} are in perfect agreement
with one loop calculations in the SM fields \cite{dp,bddms}. Thus the physical picture that emerges
is clear.
At high energies,
where particle masses can be neglected,
the effective theory is given by an unbroken $SU(2) \times U(1)$ theory for
fermions and transversely polarized gauge bosons, modified at the subleading
level by Yukawa corrections as
a unique ingredient of spontaneously broken gauge theories, and by the equivalence theorem
for longitudinally polarized gauge bosons. The contribution from soft photons
and collinear terms below the weak scale is determined by QED (including mass
terms in the corresponding logarithms).
In this contribution we will summarize the results of Refs. \cite{flmm,m1,m2} and briefly
discuss their significance at TeV energies. Including soft bremsstrahlung with a cut
on the allowed ${\mbox{\boldmath $k$}_\perp} \leq \mu_{\rm expt} \leq M$ of the emitted
real photons, and regularizing virtual IR divergences with a cutoff
${\mbox{\boldmath $k$}_\perp} \geq \mu$, we find for the semi-inclusive cross sections:
\begin{eqnarray}
&& d\sigma (p_{1}, \ldots, p_{n},g,g^\prime,\mu_{exp}) = d\sigma_{\rm Born} (p_{1},
\ldots ,p_{n},g(s),g^\prime (s))
\nonumber \\ && \times \exp \left\{ - \sum^{n_g}_{i=1} W^g_i (s,M^2)
- \sum^{n_f}_{i=1} W^f_i (s,M^2) - \sum^{n_\phi}_{i=1} W^\phi_i (s,M^2)
\right\} \nonumber \\
&&\times \exp \left[ - \sum_{i=1}^{n_f} \left( w^f_i(s,\mu^2)
- w^f_i(s,M^2) \right)
- \sum_{i=1}^{n_w} \left( w^{\rm w}_i(s,\mu^2)
- w^{\rm w}_i(s,M^2) \right) \right. \nonumber \\
&& \;\;\;\;\;\;\;\;\;\;\; \left. - \sum_{i=1}^{n_\gamma} w_i^\gamma(M^2,m_j^2)
\right]
\times \exp \left( w^\gamma_{\rm expt} (s,m_i,\mu,\mu_{\rm expt})
\right) 
\nonumber
\end{eqnarray}
The functions $W$ and $w$ correspond to the logarithmic probability to emit
a soft and/or collinear gauge boson per line, where the capital letters denote
the probability in the high energy effective theory and the lower case letter the
corresponding one from pure QED corrections below the weak scale. The matching
condition is implemented such that for $\mu=M$ only the high energy effective
solution remains.
For the contribution from scalar fields $\phi=\{\phi^\pm,\chi,H\}$ above the
scale $M$ we have
\[
W^\phi_i(s,M^2) =  \frac{ g^2(s)}{16 \pi^2} \!\! \left[ \! \left( T_i(T_i+1)+  \tan^2 \!
\theta_{\rm w}
\frac{Y^2_i}{4} \right) \!\! \left( \log^2 \!\! \frac{s}{M^2}- 4 \log \frac{s}{M^2}
\! \right) \!\!
+ \! \frac{3}{2} \frac{m^2_t}{M^2} \log \frac{s}{M^2} \right] \nonumber
\]
and for fermions:
\begin{eqnarray}
W^f_i(s,M^2) &=&  \frac{ g^2(s)}{16 \pi^2} \!\! \left[ \! \left( T_i(T_i+1)+  \tan^2 \!
\theta_{\rm w}
\frac{Y^2_i}{4} \right) \!\!
\left( \log^2 \frac{s}{M^2}- 3 \log \frac{s}{M^2}
\! \right) \right. \nonumber \\ && \left.
+ \left( \frac{1+\delta_{f,{\rm R}}}{4} \frac{m^2_f}{M^2} + \delta_{f,{\rm L}}
\frac{m^2_{f^\prime}}{4 M^2} \right)
\log \frac{s}{M^2} \right] 
\nonumber
\end{eqnarray}
and also for external transversely polarized gauge bosons: 
\begin{eqnarray}
W^g_i(s,M^2) &=& \left( \frac{\alpha(s)}{4 \pi}T_i(T_i+1)+ \frac{\alpha^\prime(s)}{4 \pi}
\left( \frac{Y_i}{2} \right)^2 \right) \log^2 \frac{s}{M^2} \nonumber \\
&&
- \left( \delta_{i,{\rm W}} \frac{\alpha(s)}{\pi} \beta_0 + \delta_{i,{\rm B}}
\frac{\alpha^\prime(s)}{\pi} \beta^\prime_0 \right) \log \frac{s}{M^2} 
\nonumber
\end{eqnarray}
with
\[
\beta_0=\frac{11}{12}C_A - \frac{1}{3}n_{gen}-\frac{1}{24}n_{h} \;\;\;,\;\;\;
\beta^\prime_0= - \frac{5}{9}n_{gen} -\frac{1}{24}n_{h} 
\nonumber
\]
where $n_{gen}$ denotes the number of fermion generations and $n_h$ the number of
Higgs doublets. Again we note that for external photon and Z-boson states we must include
the mixing appropriately as discussed in Ref. \cite{m1}.
For the terms entering from contributions below the weak scale we have for fermions:
\[
w^f_i(s,\mu^2) = \left\{ \begin{array}{lc} \frac{e_i^2}{(4 \pi)^2} \left( \log^2 \frac{s}{\mu^2}
- 3 \log \frac{s}{\mu^2} \right) & , \;\;\; m_i \ll \mu \\
\frac{e_i^2}{(4 \pi)^2} \left[ \left( \log \frac{s}{m_i^2}-1 \right) 2 \log \frac{m_i^2}{\mu^2}
\right. \\
\left.\;\;\;\;\;\;\;\;\;\;+ \log^2 \frac{s}{m_i^2} - 3 \log \frac{s}{m_i^2} \right] & , \;\;\; \mu
\ll m_i\end{array} \right.
\nonumber
\]
Analogously, for external W-bosons and photons we find:
\[
w^{\rm w}_i(s,\mu^2) =
\frac{e_i^2}{(4 \pi)^2} \left[ \left( \log \frac{s}{M^2}-1 \right)
2 \log \frac{M^2}{\mu^2}
+ \log^2 \frac{s}{M^2} \right]
\nonumber
\]
\[
w_i^\gamma(M^2,\mu^2) = \left\{ \begin{array}{lc}
\frac{1}{3} \sum_{j=1}^{n_f} \frac{e_j^2}{4 \pi^2} N^j_C
\log \frac{M^2}{\mu^2} & , \;\;\; m_j \ll \mu \\
\frac{1}{3} \sum_{j=1}^{n_f} \frac{e_j^2}{4 \pi^2} N^j_C \log \frac{M^2}{m_j^2}
& , \;\;\; \mu \ll m_j\end{array} \right.
\nonumber
\]
for the virtual corrections and for real photon emission we have in the soft
photon approximation:
\begin{eqnarray}
w_{\rm expt}^{\gamma }(s,m_i,\mu,\mu_{\rm expt})
\!\!&=&\!\!\! \left\{ \begin{array}{lc}
\sum_{i=1}^n \frac{e_i^2}{(4 \pi)^2} \left[
- \log^2 \frac{s}{\mu^2_{\rm expt}}
+ \log^2 \frac{s}{\mu^2}- 3 \log \frac{s}{\mu^2} \right]
& , m_i \ll \mu \\
\sum_{i=1}^n \frac{e_i^2}{(4 \pi)^2} \left[ \left( \log
\frac{s}{m_i^2} -  1 \right)
2 \log \frac{m_i^2}{\mu^2} + \log^2 \frac{s}{m_i^2}
\right. \\ \left. - 2 \log \frac{s}{\mu^2_{\rm expt}} \left( \log \frac{s}{m_i^2} -
1 \right) \right]
& ,
\mu \ll m_i \end{array} \right. \nonumber \\ &&
\nonumber
\end{eqnarray}
where $n$ is the number of external lines
and the upper case applies only to fermions since for $W^\pm$
we have $\mu < M$. Note that in all contributions from the regime $\mu<M$ we have
kept mass terms inside the logarithms. This approach is valid in the entire Standard
Model up to terms of order ${\cal O} \left( \log \frac{m_t}{M} \right)$.
The overall $\mu$-dependence in the semi-inclusive cross section cancels and we only have
a dependence on the parameter $\mu_{\rm expt}$ related to the experimental energy resolution.
All universal Sudakov corrections at DL and SL level exponentiate. 
The size of the DL correction at the two loop level is typically ${\cal O}(0.2 \%)$ at 1 TeV
per line on the level of the cross section and the subleading universal corrections can
be of the same order. The two loop effects are thus non-negligible. 
For processes with Yukawa enhanced SL contributions, the overall
Sudakov suppression is enhanced.
For processes
with a large angle dependence, however, we must also include terms like $\log \frac{u}{t}
\log \frac{s}{M^2}$ for which only one loop results are known \cite{dp,kps}. Also
higher order renormalization group corrections of ${\cal O} \left( \alpha^n \beta_0 \log^{2n-1}
\frac{s}{M^2} \right)$ should be inlcuded via a running coupling in each loop
correction in analogy to QCD \cite{ms2,mrg}.


\begin{references}
\bibitem{ccc} M.~Ciafaloni, P.~Ciafaloni, D.~Comelli; Phys.~Rev.~Lett. {\bf 84}:4810, 2000.

\bibitem{flmm} V.S.~Fadin, L.N.~Lipatov, A.D.~Martin, M.~Melles;
Phys.Rev. {\bf D61} (2000) 094002.

\bibitem{kl}  R. Kirschner, L. N. Lipatov; JETP {\bf 83} (1982) 488;
Phys. Rev. {\bf D26} (1982) 1202.

\bibitem{vg}  V. N. Gribov, Yad. Fiz. 5 (1967) 399 (Sov. J. Nucl. Phys. {\bf 5}
(1967) 280); Sov. J. Nucl. Phys. {\bf
12} (1971) 543; 
L. N. Lipatov, Nucl. Phys. {\bf B307} (1988) 705; \newline
V. Del Duca, Nucl. Phys. {\bf B345} (1990) 369.

\bibitem{m1} M.~Melles; hep-ph/0004056, accepted for publication in
Phys. Rev. {\bf D}.

\bibitem{mgap} M.~Melles, hep-ph/0006077 accepted for publication in
Phys. Lett. {\bf B}.

\bibitem{bw} W.~Beenakker, A.~Werthenbach; Phys. Lett. {\bf B489} (2000) 148.

\bibitem{hkk} M.~Hori, H.~Kawamura, J.~Kodaira; HUPD-0003, hep-ph/0007329.

\bibitem{m2} M.~Melles; hep-ph/0012157, submitted to
Phys. Rev. {\bf D}.

\bibitem{dp} A.~Denner, S.~Pozzorini; PSI-PR-00-15, hep-ph/0010201.

\bibitem{bddms} W.~Beenakker, A.~Denner, S.~Dittmaier, R.~Mertig, T.~Sack; Nucl. Phys. {\bf B410}
(1993) 245.

\bibitem{kps} J.H.~K\"uhn, A.A.~Penin, V.A.~Smirnov; Eur.~Phys.~J.~{\bf C17} (2000) 97.

\bibitem{ms2} M.~Melles, W.J.~Stirling; Nucl.~Phys. {\bf 564} (2000) 325.

\bibitem{mrg} M.~Melles, Phys.~Rev.~{\bf D60} (1999) 075009. 

\end{references}
\end{document}